# ACCELERATOR PHYSICS AND TECHNOLOGY RESEARCH TOWARD FUTURE MULTI-MW PROTON ACCELERATORS *

V. Shiltsev[#], P. Hurh, A. Romanenko, A. Valishev, R. Zwaska, Fermilab, Batavia, IL 60510, USA

*Abstract*

Recent P5 report [1] indicated the accelerator-based neutrino and rare decay physics research as a centrepiece of the US domestic HEP program. Operation, upgrade and development of the accelerators for the near-term and longer-term particle physics program at the Intensity Frontier face formidable challenges. Here we discuss accelerator physics and technology research toward future multi-MW proton accelerators.

## MULTI-MW ACCELERATORS: ISSUES

The 2014 Particle Physics Project and Prioritization Panel (P5) provided an updated strategic plan for the US HEP program necessary to realize a twenty-year global vision for the field. The near-term program of HEP research at the Intensity Frontier continuing throughout this decade includes the long-baseline neutrino experiments and a muon program focused on precision/rare processes. It requires: a) double the beam power capability of the Booster; b) double the beam power capability of the Main Injector; and c) build-out the muon campus infrastructure and capability based on the 8 GeV proton source. The long-term needs of the Intensity Frontier community are expected to be based on the following experiments: a) long-baseline neutrino experiments to unravel neutrino sector, CP-violation, etc.; and b) rare and precision measurements of muons, kaons, neutrons to probe mass-scales beyond LHC.

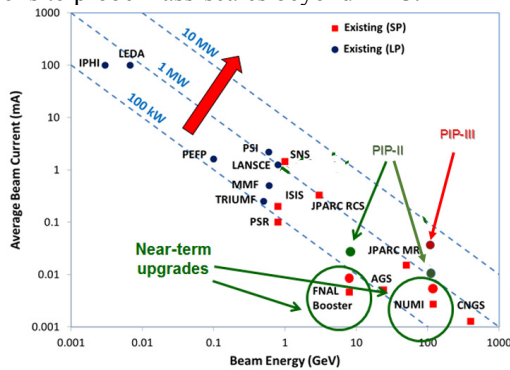

Figure 1: Accelerator beam power landscape.

Construction of the PIP-II SRF 800 MeV linac [2] is expected to address the near-term challenges. PIP-II will increase the Booster per pulse intensity by 50% and allow delivery 1.2 MW of the 120 GeV beam power from the Fermilab's Main Injector, with power approaching 1 MW at energies as low as 60 GeV, at the start of DUNE/LBNF operations ca 2023. It will also support the current 8 GeV program, including Mu2e, g-2, and the suite of short-baseline neutrino experiments; provide upgrade path for Mu2e and a platform for extension of beam power to DUNE/LBNF to multi-MW levels.

The P5 report sets longer-term sensitivity goals for the US long-baseline neutrino program. Those goals require an exposure of 600 kT*MW*yr (the product of the detector mass, beam power on target and exposure time). PIP-II offers a platform for the first 100 kt*MW*yr as in Table 1.

Table 1: Neutrino Physics Program Requirements

| | PIP-II | Beyond PIP-II (mid-term) | | |
|---|---|---|---|---|
| | 1st 10 years | 2nd 10 years | | |
| To Achieve : | 100 kT-MW-year | 500 kT-MW-year | | |
| We combine : | | Option 1 | Option 2 | Option 3 |
| Mass | 10 kT | 50 kT | 20 kT | 10 kT |
| Power | 1 MW | 1 MW | 2.5 MW | 5 MW |

The mid-term strategy towards an additional 500 kT*MW*yr after PIP-II depends on the technical feasibility of each option (see the Table) and the analysis of costs/kiloton of detector versus costs/MW of the beam power on target. To make an informed choice, extensive medium-term R&D on the effective control of beam losses in significantly higher current proton machines and on multi-MW targetry is needed [3].

There are two approaches for the multi-MW proton machine (currently tagged as PIP-III, see Fig. 1) – rapid cycling synchrotron and SRF linac. Attainment of the required beam intensities in synchrotrons is only possible with greatly reduced particle losses stemming from space-charge forces and coherent and incoherent beam instabilities. Modern SRF proton linacs can accelerate the needed currents but their cost/performance ratio needs to be significantly reduced compared to, e.g., the Project X facility [4] to be financially feasible. For both avenues, high-power targetry technology needs to be considerably enhanced to contribute to the cost and feasibility of any multi-MW superbeam facility.

In 2014-15, the HEPAP subpanel has developed *"A Strategic Plan for Accelerator R&D in the US"* [5] which recommends three R&D activities toward next step intensity frontier facility:

a) experimental studies of novel techniques to control beam instabilities and particle losses, such as integrable beam optics and space-charge compensation at IOTA ring at Fermilab;

b) exploration of the SRF capital and operating cost reductions through transformational R&D on high-Q cavities and innovative materials such as Nb-Cu composites, Nb films and $Nb_3Sn$; cavity performance upgrades through novel shapes and field emission elimination;

c) understanding the issues in multi-MW beam targets and developing mitigation techniques, new technologies and new designs.

## EXPERIMENTAL R&D AT IOTA RING

Progress of the Intensity Frontier accelerator based HEP is hindered by fundamental beam physics phenomena such as space-charge effects, beam halo formation, particle losses, transverse and longitudinal instabilities, beam loading, inefficiencies of beam injection and extraction, etc. The IOTA facility at Fermilab [6, 7] is being built as a unique test-bed for transformational R&D towards the next generation high-intensity proton facilities – see Fig. 2. The experimental accelerator R&D at the IOTA ring with protons and electrons, augmented with corresponding modeling and design efforts will lay foundation for novel design concepts, which will allow substantial increase of the proton flux available for HEP research with Fermilab accelerators to multi-MW beam power levels at very low cost. The IOTA facility will also become the focal point of a collaboration of universities, National and international partners.

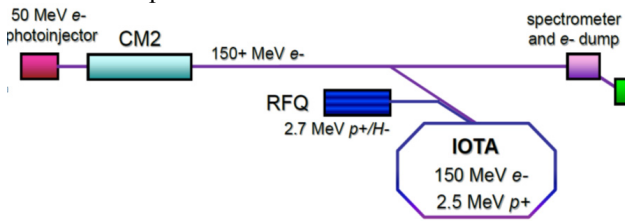

Figure 2: Schematic layout of the IOTA facility at FNAL.

The goal of the IOTA research program is to carry out experimental studies of transformative techniques to control proton beam instabilities and losses, such as *integrable optics* [8] with non-linear magnets and with electron lenses, and *space-charge compensation* with electron lenses and electron columns [9, 10] at beam intensities and brightness 3-4 times the current operational limits, i.e., at the space-charge parameter $\Delta Q_{SC}$ approaching or even exceeding 1. Several experiments are planned at IOTA – see Fig. 3:

**Integrable Optics Tests with Electrons** have goals to create IO accelerator lattice with several additional integrals of motion (angular momentum and McMillan-type integrals, quadratic in momentum), test "reduced integrability" with octupoles, confirm the IO dynamics with pencil e- beam, and confirm the particle stability over tune spreads ~0.5/cell, including possible crossing of integer resonances. The tests call for employment of narrow (pencil) pencil e- beam to assure 1% or better measurement and control of beta-functions, and 0.001 or better control of the betatron phases.

**IO with Non-linear Magnets, Test with Protons** will demonstrate nonlinear integrable optics with protons with a large betatron frequency spread $\Delta Q_{SC}$>1 and stable particle motion in a realistic accelerator design. One would expect to observe greatly reduced space-charge losses, acceptable stability to perturbations in 3D, stable coherent and incoherent dynamics. It will require advanced diagnostics, e.g., 1% or better measurement and control of beam distribution moments, halo and losses.

**IO with e-lens(es), Tests with Protons** to demonstrate IO with non-Laplacian electron lenses with protons with a large betatron frequency spread $\Delta Q_{SC}$>1 and stable particle motion in a realistic accelerator design. Sensitivity to deviations of the electron charge distribution from the ideal one $n(r)=1/(1+r^2)^2$ will be studied, too.

**Space-Charge Compensation (SCC) with e-lens(es), Test with Protons** has the main goal of demonstrating SCC with Gaussian ELs with protons with a large betatron frequency spread $\Delta Q$>0.5 and stable particle motion in a realistic accelerator design.

Similar *SCC tests* are envisioned *with electron columns*.

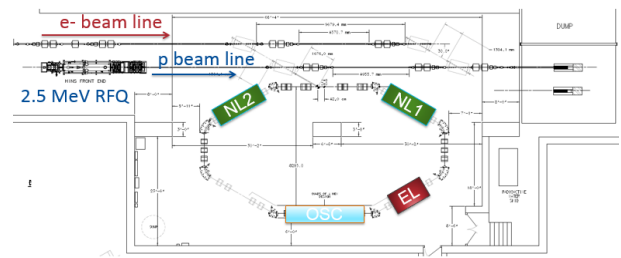

Figure 3: IOTA ring, its electron and proton injection lines and experimental areas.

## COST EFFECTIVE SRF TECHNOLOGY

Superconducting RF is the state-of-the-art technology for a majority of near-, mid- and far-term accelerators due to its unmatched capability to provide up to 100% duty factor and large apertures to preserve the beam quality. The very successful SRF R&D in the past has been predominantly focused on improving gradients, extending from 3 MV/m to >35 MV/m. Recent shift of focus towards decreasing costs has led to several major breakthroughs: 1) nitrogen doping for ultra-high Q cavities, which opens up more than a factor of two savings in cryogenics capital and operational costs [11]; 2) Nb/Cu composite material and monolithic techniques of cavity manufacturing; these avenues promise a factor of >2 reduction in cavity material and manufacturing costs with performance comparable to bulk Nb cavities; 3) $Nb_3Sn$ cavities for 4.2K operation.

Three different routes to drastically lower the required power capacity and therefore capital and operational costs of such accelerators are to be explored: a) *Capital and Operating Cost Reduction by Raising the Q*. Recently it was discovered at FNAL that quality factors (Q) of bulk niobium cavities can be drastically increased by factors of 2-4 using the nitrogen doping procedure [11]. Furthermore, changing the cool-down procedure around the critical temperature 9.2K was discovered as the extremely efficient way to minimize the trapped magnetic flux contribution to rf losses, and thus to preserve the ultralow surface resistance in the cryomodule environment [12]. We will establish optimal parameters of doping and cool-down for the high Q Nb technology for different frequencies and operating conditions and also explore the underlying physics of both doping and the cool-down effects; b) *Raising operating temperature.* Increasing operating temperature of superconducting accelerator cavities to > 4.2K promises a dramatic increase in the cryoplant efficiency. Furthermore, cryogen-free operation may become possible for smaller-scale accelerators using economical cryo-coolers. The proof-of-principle exists that Nb3Sn cavities provide the same quality factors at >4.2K as bulk niobium cavities do at 2K [13]. We will develop Nb3Sn cavities with the vapor diffusion method and use the existing set of advanced cavity characterization tools to understand limitations and guide the development; c) *SRF material cost reduction:* There exists a proof-of-principle that 1.3 GHz Nb-Cu composite based spun cavities, can reach high gradients. In collaboration with Cornell University we will use the existing Nb-Cu sheets to spin the cavities at INFN or US industry (e.g. AES) to complete the 650 MHz cavities with flanges as the first step followed by scaling to 325 MHz if successful. Recent breakthrough in Nb film deposition technology allows films of unprecedented quality with the residual resistivity ratio (RRR) approaching or exceeding 200-300, which is currently the standard for bulk SRF cavities. The rf properties of these films will be tested. Given the confirmed low surface resistance on samples we then plan to proceed with 650 MHz Nb/Cu cavity prototyping followed by scaling to 325 MHz.

## HIGH POWER TARGERY R&D

Mega-watt class target facilities present many technical challenges, including: radiation damage, rapid heat removal, high thermal shock response – see, e.g., Fig. 4, highly non-linear thermo-mechanical simulation, radiation protection, and remote handling [14]. The major goal of the envisioned R&D program for the next decade is to enable well-justified design simulations of high intensity beam/matter interactions using realistic, irradiated material properties for the purposes of designing and predicting lifetimes of multi-MW neutrino and muon target components and systems. This requires: a) irradiated material properties to be measured/evaluated for relevant targetry materials over a range of temperatures (300 – 1300 K), radiation damage (0.1 – 20 DPA (Displacements Per Atom)) and relevant helium production rates (500 – 5000 atomic parts per million/DPA); b) thermal shock response to be evaluated for relevant targetry materials over a range of strain rates (100 – 10000 $s^{-1}$); c) development and validation of simulation techniques to model material response to beam over the time of exposure (accounting for accumulation of radiation damage and high spatial gradients); d) development of enabling technologies in target materials, manufacturing techniques, cooling technologies, instrumentation, radiation protection, and related systems to meet the targetry challenges of multi-MW and/or high intensity (> 500 MW/$m^3$ peak energy deposition) requirements of future target facilities.

*Radiation damage studies* include investigations of materials of high interest (currently graphite, beryllium, tungsten and titanium alloys) under the RaDIATE R&D program [15]. The most major of these activities involve Post-Irradiation Examination (PIE) of previously irradiated materials recovered from spent target components (e.g. NuMI proton beam window), low-energy ion and high energy proton irradiations at available beam facilities (e.g. Brookhaven Linac Isotope Producer [16]), and experiments designed to help correlate low energy ion irradiations to high energy proton irradiations.

*Thermal shock response studies* include in-beam thermal shock experiments of various grades of commercially pure beryllium at the HiRadMat Facility [17] at CERN (e.g. HRMT-24, "BeGrid" [18]) and high strain rate testing of candidate materials to develop strength and damage models.

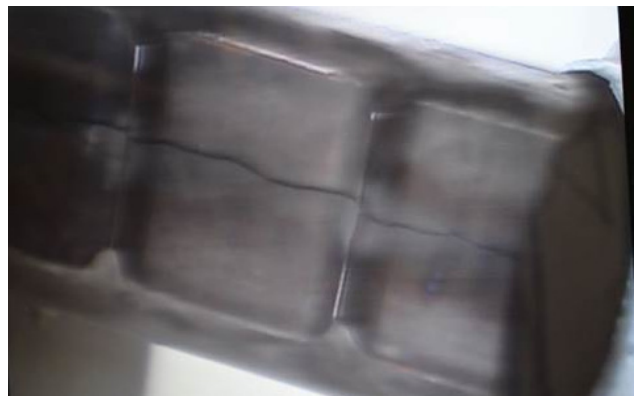

Figure 4: Graphite fins from NuMI target NT-02. Cracking is thought to be the result of material degradation and thermal shock from the beam which passes from left to right (courtesy V. Sidorov, Fermilab).

## REFERENCES

[1] *Building for Discovery* (P5 Report, May 2014); see at http://science.energy.gov/hep/hepap/reports/
[2] P. Derwent, S. Holmes, V. Lebedev, arXiv:1502.01728
[3] V. Shiltsev et al., MOPAC16, Proc. of NA-PAC'13, Pasadena, CA, USA (2013); http://www.JACoW.org
[4] S. Holmes et al., arXiv:1306.5022 (2013).
[5] See at http://science.energy.gov/hep/hepap/reports/
[6] *ASTA Facility Proposal*, Fermilab-TM-2568 (2013).
[7] A. Valishev et al., MOPMA021, these proceedings, Proc. of IPAC'15, Richmond, VA, USA (2015); http://www.JACoW.org
[8] V. Danilov, S. Nagaitsev, PRTSAB **13**, 084002 (2010).
[9] A. Burov et al., Fermilab TM-2125 (2000).
[10] V. Shiltsev, TUPMN106, Proc. of PAC'07, Albuquerque, NM, USA (2007); http://www.JACoW.org

[11] A. Grassellino et al., Superc. Sci. Tech. **26**, 102001 (2013).
[12] A. Romanenko et. al., Appl. Phys. Lett. **105**, 234103 (2014).
[13] S. Posen et al., Appl. Phys. Lett. **106**, 082601 (2015).
[14] P. Hurh et al., THPFI082, Proc. of IPAC'13, Shanghai, China (2013); http://www.JACoW.org
[15] www-radiate.fnal.gov
[16] N. Simos et al., Fermilab-Conf-10-480-APC (2010).
[17] A. Fabich et al., THPFI055, Proc. of IPAC'13, Shanghai, China (2013); http://www.JACoW.org
[18] K. Ammigan et al., "Examination of Beryllium under Intense High Energy Proton Beam at CERN's HiRadMat Facility," WEPTY015, these proceedings, Proc. of IPAC'15, Richmond, VA, USA (2015); http://www.JACoW.org